# CODE-DIVISION MULTIPLEXED RESISTIVE PULSE SENSOR NETWORKS FOR SPATIO-TEMPORAL DETECTION OF PARTICLES IN MICROFLUIDIC DEVICES

*Ningquan Wang, Ruxiu Liu, Roozbeh Khodambashi, Norh Asmare, and A. Fatih Sarioglu*
School of Electrical and Computer Engineering, Georgia Institute of Technology, Atlanta, USA

## ABSTRACT

Spatial separation of suspended particles based on contrast in their physical or chemical properties forms the basis of various biological assays performed on lab-on-a-chip devices. To electronically acquire this information, we have recently introduced a microfluidic sensing platform, called Microfluidic CODES, which combines the resistive pulse sensing with the code division multiple access in multiplexing a network of integrated electrical sensors. In this paper, we enhance the multiplexing capacity of the Microfluidic CODES by employing sensors that generate non-orthogonal code waveforms and a new decoding algorithm that combines machine learning techniques with minimum mean-squared error estimation. As a proof of principle, we fabricated a microfluidic device with a network of 10 code-multiplexed sensors and characterized it using cells suspended in phosphate buffer saline solution.

## INTRODUCTION

Integrated lab-on-a-chip systems can be realized by incorporating passive electrical sensors on microfluidic devices. Such devices enable counting, sizing and electrical characterization of small particles suspended in liquids, and can be used in applications ranging from bio-medicine to environmental monitoring [1]. Meanwhile, obtaining information on the spatio-temporal manipulation of particles under various force fields in lab-on-a-chip devices enables one to perform a wider array of biophysical or biochemical analyses [2]. However, this spatio-temporal information is typically obtained through microscopic imaging of the chip, which constitutes a bottleneck in employing these devices outside the laboratory. Therefore, a microfluidic device with integrated electrical sensors that can spatio-temporally track particles can help realize a fully integrated, low-cost system that would especially be useful for analysis of samples in mobile and/or resource-limited settings.

Recently, we have introduced a microfluidic platform with an integrated network of multiplexed electronic sensors, called Microfluidic CODES [3], [4], that combines the resistive pulse sensing [5] with the code division multiple access (CDMA) [6]. In Microfluidic CODES, each sensor in the network is formed by a distinctly patterned array of co-planar electrodes and therefore produces a signal distinguishable from other sensors' when it detects a particle. Moreover, sensor signals are specifically designed to be orthogonal Gold sequences [7], commonly used in CDMA uplink, to ensure reliable recovery even when they interfere with other sensors in the network. This allows Microfluidic CODES platform to effectively compress multidimensional information on spatio-temporal manipulation of particles in a microfluidic chip into a decodable electrical time waveform.

Here, we enhance the multiplexing capacity of the Microfluidic CODES platform by employing sensors designed to produce unipolar, non-orthogonal sequences, rather than specially designed bipolar orthogonal Gold sequences we used earlier. We also introduce a decoding algorithm that combines machine learning techniques with minimum mean-squared error (MMSE) estimation to decode sensor signals. As a proof of principle, we developed a microfluidic device with a network of 10 code-multiplexed sensors and characterized it using a suspension of cells in phosphate buffer saline (PBS) solution.

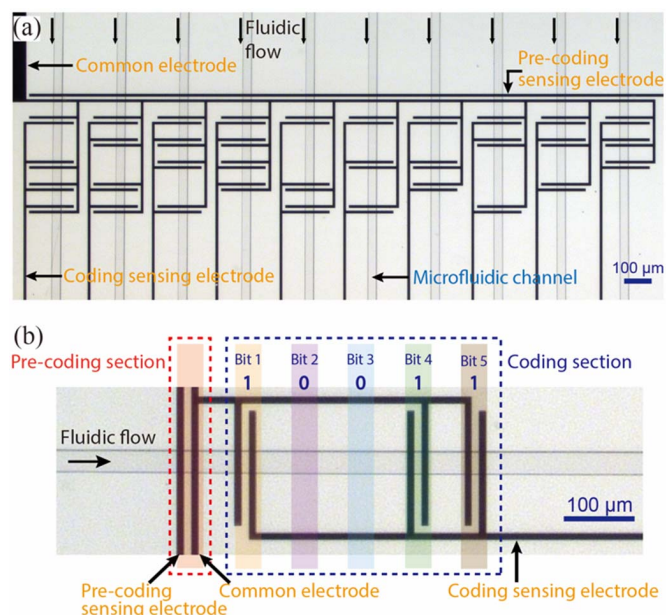

*Figure 1: (a) An image of the fabricated microfluidic device with 10 code-multiplexed sensors. Each sensor consists of a pre-coding sensor followed by a 5-bit coding section. (b) A close-up of the sensor designed to generate code "10011".*

## DEVICE DESIGN

Our code-multiplexed sensor network is formed by three micromachined coplanar surface electrodes aligned with microfluidic channels (Figure 1a). Each sensor in the network is made up of multiple pairs of 10 μm-wide electrode fingers arranged in a distinct pattern. Across the sensor network, the current flow is confined to sections of the microfluidic channels that fall in between different electrodes. Therefore, only when particles occupy these locations, the sensor output is modulated, an event we interpret as a positive bit (i.e., "1") in a code signal (Figure 1b). In operation, particles flowing in microfluidic channels sequentially occupy these regions, producing a unipolar pulse sequence effectively encoding desired locations on the microfluidic device with a distinct digital label.

The main advantage of using non-orthogonal codes in code-multiplexing sensors is that a larger set of distinct codes is available to assign to sensors for a given code length (i.e., the number of bits for a digital code). This allows us to construct more compact sensors to generate shorter codes reducing the total sensing volume. As a result, fewer particles overlap in the sensing volume for a given sample density and less interference between individual sensors improves the decoding accuracy, and hence, the multiplexing capacity. On the other hand, decoding of interfering non-orthogonal sensor signals suffers from higher crosstalk between signals, which we mitigate by optimized sensor and decoding algorithm design.

To aid the decoding of interfering non-orthogonal code signals, our device employs a pre-coding sensor placed adjacent to the coding electrodes. The pre-coding sensor consists of a single pair of electrodes. One of these electrodes is a common electrode shared with the coding section. The pre-coding sensor extends into all sensors in the network and its output is separate from coding electrodes. In operation, each particle flowing in a microfluidic channel generates a single unipolar pulse in the pre-coding sensor signal followed by a digital code signal generated by the coding electrodes. Because the pre-coding sensor has a smaller footprint than the coding sensor, there is a lower probability of interference between pre-coding sensor signals due to overlapping particles. From each pulse in the pre-coding sensor signal, we determine the size, speed and relative timing of the corresponding particle and use these parameters to reduce the error rate and the computational complexity in decoding interfering code signals. Once decoded, code signals are then used to obtain the particle's location (i.e., the specific sensor that it interacted with).

## FABRICATION

Our microfluidic device consists of glass substrate with patterned surface electrodes and a polydimethylsiloxane (PDMS) microfluidic layer. We fabricated surface electrodes using a lift-off process. 1.5 µm -thick negative photoresist (NR9-1500PY, Futurrex) was spun on a 4-inch borosilicate glass wafer and then patterned using conventional photolithography to define the sensor network layout. 100 nm-thick Cr/Au film stack was deposited on the wafer using e-beam evaporation followed by the etching of the underlying photoresist in acetone under sonication. Finished wafer was then diced into chips using a wafer saw. Next, we fabricated the microfluidic layer using soft lithography. To create the mold, we coated a 4-inch silicon wafer with 15 µm-thick SU-8 photoresist (SU8-2015, MicroChem) and patterned the SU-8 film using photolithography. The mold was then coated with PDMS prepolymer and cross-linker pre-mixed at a 10:1 ratio. Polymer film was first degassed in vacuum and subsequently cured in an oven at 65 °C for 4 hours. Cured PDMS was peeled off from the mold, and cut into small pieces using a scalpel. To finalize the device, both the PDMS part and the glass chip containing the surface electrodes were first activated in an oxygen plasma, aligned and then bonded under a microscope.

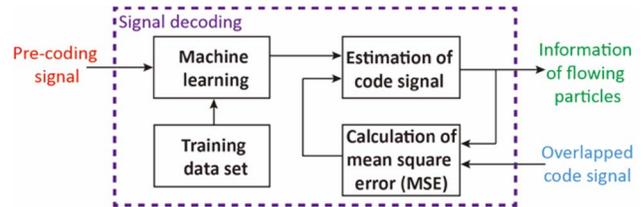

*Figure 2: Block diagram of the decoding algorithm.*

## DECODING ALGORITHM

Our decoding algorithm uses the pre-coding sensor signal and a training dataset to estimate parameters such as the number of concurrent particles in a certain time window, size and speed of each particle, and relative timing of particles, and then generates all probable combinations based on a pre-generated template library to estimate the signal with MMSE (Figure 2). It should be noted that the information from the pre-coding sensor significantly reduces the parameter space for estimation in this process. Moreover, the differences in particle sizes and speeds ensure a unique solution to the estimation problem, while such variations are detrimental to the performance of orthogonal code based multiplexing.

A critical step in our decoding algorithm is accurate estimation of potential code waveforms for a particle from the corresponding pulse in the pre-coding sensor signal. To achieve this, we use the pulse from the pre-coding sensor signal as a template to estimate the pulses in the code waveform for the same particle. However, pulse amplitudes are not the same between different electrode pairs even within the same sensor and their ratio depends on the device-specific fabrication parameters as well as circuit level effects such as coupling between different electrode pairs. To overcome this problem, we analyze a set of recorded signals for each sensor and calculate average peak amplitudes to determine scaling factors for each pulse in the code signal relative to the pre-coding sensor pulse.

Another important parameter to be determined in the estimation of the code signal is its duration, which sets the time delay between pulses in the code signal. To estimate the signal duration, we need to calculate the particle flow speed from its single pulse in the pre-coding sensor signal. For this purpose, we employed a machine learning technique, called *K*-nearest neighbors (*K*-NN) [8]. In this process, first a training dataset is generated from a set of pre-coding signal pulses for non-overlapping particles and their corresponding flow speeds. The particle flow speeds are determined by calculating the time difference between the peaks of the pre-coding sensor pulse and the pulse corresponding to the first positive bit in code signal. We extract four features from each pre-coding signal pulse in the dataset: $X_1$, $X_2$, $X_3$, and $X_4$ representing the peak amplitude, and the full width (time) at ¾, ½, and ¼ of the peak amplitude, respectively. Based on these, for every pulse in the pre-coding sensor signal, we identify *K* most resembling (i.e., minimum Euclidian distance) pre-coding sensor pulses in the training dataset - also known as its nearest neighbors. The flow speed of the particle is then estimated by averaging the speed values of these nearest neighbors. Finally, assuming the particle speed remains constant within the sensor, we estimate the duration of the code signal and construct the code waveform template for the particle.

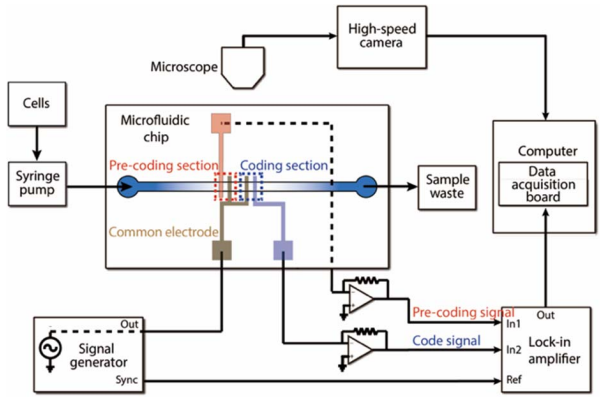

*Figure 3: Schematic drawing of the experimental setup.*

Once the template code waveforms for all sensors are established for every particle, we combine estimated code templates for particles with specific time delays set according to the pre-coding sensor data. In this process, we consider all possible combinations of sensors in the network and choose the waveform with MMSE as the decoder output.

## EXPERIMENTAL RESULTS

To test our device and decoding algorithm, we used live cells as suspended particles. Human cancer cells, propagated in a $CO_2$ incubator, were spiked in a PBS solution to prepare the test sample. The sample was loaded into a syringe and driven through the microfluidic device using a syringe pump. To electrically detect the flow of cells, the code-multiplexed sensor network was excited by applying a 500 kHz sine wave to the common electrode. This signal frequency was chosen to be high enough so as to not be affected by the double layer effect at the electrode-electrolyte interface, and low enough to avoid the Maxwell−Wagner dispersion [9]. Current signals from the pre-coding and coding sensors were converted to voltage signals using two transimpedance amplifiers, and their RMS values were measured by a two-channel lock-in amplifier (Zurich Instruments HF2LI). Lock-in amplifier outputs were sampled at 50 kHz into a computer through a data acquisition board. Recorded sensor signals were processed using custom software written in LabVIEW and MATLAB.

The cell flow in our device was simultaneously recorded using an inverted optical microscope (Nikon Eclipse Ti) equipped with a high-speed camera (Vision Research Phantom v7.3) at 1000 frames per second. The optical data were used to evaluate the performance of our device and the decoding algorithm. A schematic of our measurement system is shown in Figure 3.

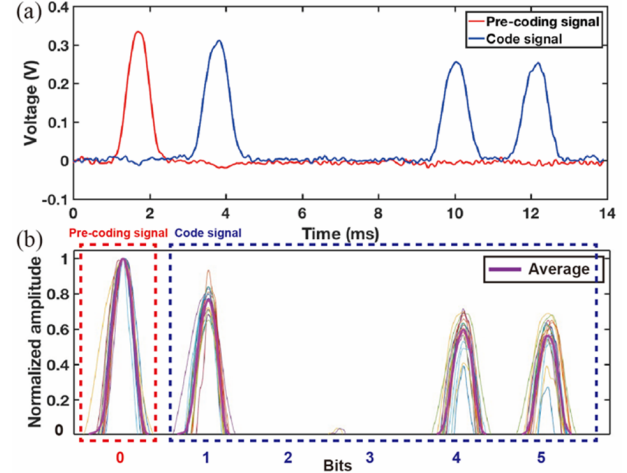

*Figure 4: (a) Recorded signals from the sensor designed to produce the digital code 10011 when it detects a cell. (b) Averaging of normalized code pulses from different cells to determine peak amplitude ratios between different pulses.*

Figure 4a shows recorded pre-coding sensor signal and the code signal (10011) corresponding to the sensor shown in Figure 1b together. Note that, in the sensor layout, the distance between the electrode pairs associated with the pre-coding sensor and the first positive bit is the same with the distance between the electrode pairs encoding the last two positive bits. Therefore, similar time delay between these pulses indicate a constant flow speed for the cell as it traverses the sensor electrodes.

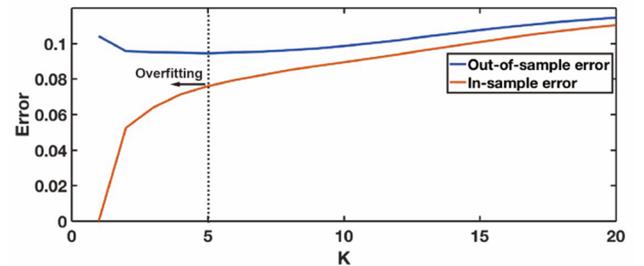

*Figure 5: Calculation of the optimum K to be used in the K-NN algorithm for a 400-cell training dataset.*

Our results also demonstrate that pulses in the pre-coding sensor signal and the code signal have different peak amplitudes. We compared signals from the same sensor for a large number of cells and observed a similar amplitude pattern among normalized signals (Figure 4b). Based on this observation, we averaged amplitude- and time-normalized pulses recorded from different cells to determine peak amplitude ratios between the pre-coding sensor pulse and the code signal pulses for each sensor. We recorded these values as scaling factors that we use to construct code signal templates in our decoding algorithm as explained before.

The selection of *K* is crucial in the *K*-NN algorithm used as part of our machine learning approach to determine the cell flow speed from the pre-coding sensor signal. To find the optimum *K*, we used the repeated random sub-sampling validation on our training dataset consisting of recorded signals from 400 cells. We randomly split the original training dataset into two equally populous subsets; a training and a test subset. The first evaluation was done solely within the training subset (in-sample evaluation). Each sample in the training subset was treated as a test sample. The second evaluation was done between the training subset and the test subset (out-of-sample evaluation). In this case, each sample in the test subset was predicted by the samples in the training subset and the error was evaluated. Finally, the random split and error evaluation process was performed 50 times for different *K* values, and all the in-sample error and out-of-sample error values were averaged (Figure 5). We determined that the optimum *K* with minimum out-of-sample error for our dataset equals 5.

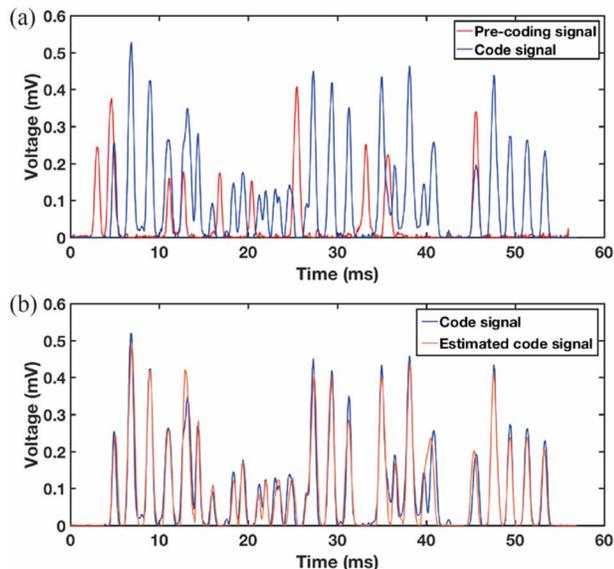

*Figure 6: Decoding of the signal corresponding to 10 cells.*

Employing optimized parameters, we processed interfering code signals from cells using our decoding algorithm. Figure 6a shows a recorded pre-coding sensor signal and the code signal from 10 cells flowed over the sensor network in a duration of 60 ms. In the decoding process, 10 pre-coding sensor pulses were first analyzed by the machine learning stage to estimate the speed of each cell. Based on these estimations, each pre-coding pulse was used to generate 10 possible code signals based on the templates (i.e., pulse scaling factors) created for the 10 sensors in the network. All possible combinations of these estimated code signals for 10 pre-coding sensor pulses were analyzed and the combination yielding the MMSE fit to the recorded code signal was identified as the optimal solution. The output from our decoding algorithm and the recorded code signal closely match (Figure 6b). Our result indicates that 10 cells flowed through different microfluidic channels in the following order: 8, 7, 8, 6, 6, 7, 8, 7, 9, 7 (channels numbered from right to left in Figure 1). We validated this result using the high-speed optical microscopy images.

## CONCLUSION
We have demonstrated a Microfluidic CODES platform based on non-orthogonal codes and also a machine-learning based algorithm for decoding sensor signals. Our results from the analysis of cells suspended in solution show that particles spatially distributed in a microfluidic device can successfully be tracked by our system through direct electrical sensing.

## ACKNOWLEDGEMENTS

This work was supported by National Science Foundation Award No. ECCS 1610995.

## CONTACT

*A. F. Sarioglu, tel: +1-650-7400764; sarioglu@gatech.edu